\documentstyle[12pt]{article}
\newcommand{\be}{\begin{equation}}
\newcommand{\bea}{\begin{eqnarray}}
\newcommand{\eea}{\end{eqnarray}}
\newcommand{\ba}{\begin{array}}
\newcommand{\ea}{\end{array}}
\newcommand{\ee}{\end{equation}}

\expandafter\ifx\csname mathbbm\endcsname\relax

\else

\fi \textheight 22cm \textwidth 15cm \topmargin 1mm
\def\IR{\relax{\rm I\kern-.18em R}}

\begin{document}
\begin{titlepage}
\begin{center}
\vspace*{15mm}
{\Large {\bf Cosmological Solution from D-brane motion in NS5-Branes background }\\ }
\vspace*{25mm}
\vspace*{1mm}
{Hossein Yavartanoo  \footnote{yavar@ipm.ir}}
 \\
\vspace*{1cm}
{\it Institute for Studies in Theoretical Physics
and Mathematics (IPM)\\
P.O. Box 19395-5531, Tehran, Iran \\ \vspace{3mm}}
\vspace*{2cm}
\end{center}
\begin{abstract}
We study dynamics of a D3-brane propagating in the vicinity of k
coincident NS5 branes. We show that when $g_s$ is small, there
exists a regime in which dynamics of the D-brane is governed by
Dirac-Born-Infeld action while higher order derivative in the
expansion can not be neglected. This leads to a
restriction on how fast scalar field may roll. We analyze the motion
of a rolling scalar field in this regime, and extend
the analysis to cosmological systems obtained by coupling this
type of field theory to four dimensional gravity. It also leads to
some FRW cosmologies, some of which are related to those obtained with
tachyon matter.
\end{abstract}
\end{titlepage}
\section{Introduction}
\par
The dynamics of branes in string theory is an important subject,
which we hope to provide clues about the fundamental description
of the string theory. The time dependent nature of the solution is
challenging and includes many puzzling issues, while important for
many applications in cosmological problems. One of the most well
studied subject is the time evolution of unstable D-branes in flat
backgrounds, that is the problem of rolling tachyons
\cite{tachyon}-\cite{tachyonm5}(see also
\cite{GutperleAI,LarsenWC}).
\\ \par
In studies of D-branes in string theory  one often finds open
string tachyon in the lowest lying open string states. This
happens in bosonic and superstring theory as well, which admit
time dependent solutions describing the rolling of the tachyon
away from the maximum of the potential.
\\
\par
The rolling of tachyon towards a minimum of the potential in some
cases can be understood exactly in the context of classical open
string theory. Thus, in general the study of tachyon condensation
provides useful insights on the time-dependent solutions in string
theory. These solutions involve the tachyon field, in which the
tachyon rolls away from the maximum of the potential, also display
novel features. They are interesting solutions, both in describing
the early universe and in studying various dynamical issues.
\\ \par
A more ambitious problem may be time dependent string or brane
dynamics in curved backgrounds. A significant example is studying
instability of (BPS) D-brane, associated with the presence of
NS5-brane in its vicinity. Recently Kutasov, using the
Dirac-Born-Infeld action, has studied the time evolution of
D-branes near the stack of NS5-branes(for earlier works see also
\cite{ElitzurPQ}), emphasizing the formal resemblance to the
rolling tachyon problem. As it was shown this instability is
closely related to the tachyon condensation problem and one can
say a lot about features of time dependent solutions on it.
\\ \par
Indeed a configuration of parallel NS5-branes and D3-brane breaks
supersymmetry completely, since the NS5-brane and  D-brane
preserve different halves of the supersymmetry of type II string
theory. The D-brane carries RR charge, but this charge can leak
from D-brane to the NS5-brane. If we place D-brane at a finite
distance from a stack of NS5-branes, it will experience an
attractive force and start moving towards the fivebranes. In this
paper we will studying some cosmological consequences which arise
from D-brane rolling down the throats of NS5-branes.
\\ \par
There are several cosmological models which are based on the
dynamics of unstable branes in string theory (see $e. g. $
\cite{dflation}-\cite{jarv} and references therein). Brane
cosmology assumes that our universe starts out with branes
embedded in higher dimensional spacetime, either stable or
unstable. In these scenarios a natural candidate for inflaton is
the brane mode, whose expectation value describes the inter brane
separation. The dynamics of inflation is therefore governed by the
interaction between D-branes. The potential which gives rise to
the inflation during the slow roll epoch is well studied in string
theory when the branes are far apart. When the brane are
sufficiently close to each other, tachyons develop. The slow
change in inter-brane separation is followed by quick roll down
the tachyonic direction.
\\
\par
In \cite{Dccel} a ``D-ecelerations" mechanism for slowing scalar
field motion was identified in the context of strongly 't Hooft
coupled conformal field theories. The slow motion of the scaler is
understood in simple way from gravity side of the AdS/CFT
correspondence which provides  the effective description of the
system. In terms of the effective action this behavior occurs due
to the higher derivative terms encoded in Dirac-Born-Infeld (DBI)
action. When the effective field theory is subsequently coupled to
gravity, this mechanism leads to inflation in much steeper
potential than are allowed in standard weakly coupled slow roll
inflation.
\\ \par
In this paper, we use this idea to study cosmological solution
arising from moving a D-brane in the vicinity of  NS5-branes. We
show that there exists a regime, in which the string coupling at
the location of the D-brane is small and then we can use DBI
action to study the dynamics  of D-brane in this background while
higher derivative terms in DBI action can not be neglected. The
non-analytic behavior of the square-root of DBI action in this
regime gives rise to a speed limit restriction on how fast the
scalar field on the brane (inflaton) can roll. We study this
system when the effective field theory is coupled to gravity and
we show that this mechanism leads to inflation for some
potentials.
\\ \par
The paper is organized as follows. We start in section 2 by
reviewing the near horizon geometry on NS5-branes (the CHS geometry)
and DBI action for D-brane in this background.
In section 3, we couple this system to gravity and study
cosmologies arising from our low energy effective action. In
section 5 we try to solve the equation of motion for scale factor
for some different potentials, and we show that there exists
an inflationary solution in some cases.
\newpage
\section{D-brane effective action in the NS5-brane background}
Consider a stack of k coincident NS5-branes in type II string
theory, stretched in directions $(x^1, ..., x^5)$ and localized in
$\vec{x} = ( x^6, x^7, x^8, x^9)$. The direction along the
worldvolume of the fivebranes will be denoted by $x^{\mu}, \mu =
0, 1, 2 ,... 5$; those transverse to the branes will be labelled
by $x^m, m=6, 7, 8, 9$.  The background fields are given by CHS
solution \cite{CallanAT}. \\ \par The metric, dilaton and NS
B-field are \bea \label{NS}
ds^2&=&dx_{\mu}dx^{\mu}+H(r)(\;dr^2+r^2d\Omega_3^2\;) \cr\cr
e^{2\Phi}&=&g_s^2H(r) \cr\cr H_{ijk}&=&
-\epsilon_{ijk}^l\;\partial_l \Phi \eea Here  $r$ is the radial
direction perpendicular to NS5-branes and $d\Omega_3^2$ is the
metric on $S^3$. $H_{ijk}$ is the field strength of the $NS$
B-field and $H (r)$ is a harmonic function on the transverse space
\be H(r)=1+{k \alpha' \over r^2}. \ee We are interested is the
dynamics of BPS Dp-brane in this background. The D-brane is
parallel to the fivebrane, $i. e.$ it is extended in some or all
of the fivebrane worldvolume directions $x^{\mu}$ and point like
in the directions transverse to the fivebranes $(x^6, x^7, x^8,
x^9)$. Without loss of generality, we can take the worldvolume of
the Dp-brane to fill the directions $(x^0, x^1, ..., x^p)$. We
will label worldvolume of the Dp-brane by $x^{\mu}$ as well, but
one should bear in mind that here the index $\mu$ only runs over
$\mu= 0,1,2, ... p$, with $p \leq 5$. \\ \par Although the D-brane
in question is BPS, in the presence of the fivebranes it is
unstable. Indeed, a configuration of parallel NS5-branes and
Dp-brane breaks supersymmetry  completely, since the fivebrane and
D-brane preserve different halves of the supersymmetry of type II
string theory. The Dp-brane carries RR charge, but this charge can
leak from D-brane to fivebranes.
\\ \par
If we place a BPS Dp-brane at finite distance from a stack of
NS5-branes, it will experience an attractive force, and thus
start moving towards the fivebranes. The mass per unit volume of a
D-brane in string units is $1/g_s$, where $g_s$ is the string
coupling. While that of a NS5 brane goes like $1/g_s^2$.
Therefore at weak string  coupling NS5 branes are much heavier
than D-branes. We can take the NS5 brane to be static and  study the motion of a D-brane
in this background. \\
\par Consider a Dp-brane in this background, we label the
worldvolume of a D-brane by $\xi^{\mu}$, $\mu= 0,1,2,...,p$, and
use  reparametrization invariance on the world-volume of the
D-brane to set $\xi^{\mu}=x^{\mu}$. The position of the D-brane in
the transverse directions $(x^6,..,x^9)$ gives rise to scaler
fields on its world-volume, $(X^6(\xi_{\mu}), ...,
X^9(\xi_{\mu}))$. The effective action on the world-volume on the
D-brane is the DBI action \cite{TseytlinDJ} \be \label{dbi1}
S=-T_p \int d^{p+1} \xi e^{-(\Phi-\Phi_0)}\sqrt{-{\mathrm
det}(G_{\mu \nu} + B_{\mu\nu})}\;. \ee $T_p$ is the tension of the
Dp-brane and $G_{\mu \nu}$, $B_{\mu\nu}$ are induced metric and
B-field on the D-brane. They are related to the metric and B-field
in ten dimension by \bea G_{\mu\nu}&=& \frac{\partial
X^A}{\partial \xi^{\mu}} \frac{\partial X^B}{\partial \xi^{\nu}}
G_{AB}(x)\;, \cr\cr B_{\mu\nu}&=& \frac{\partial X^A}{\partial
\xi^{\mu}} \frac{\partial X^B}{\partial \xi^{\nu}} B_{AB}(x)\;.
\eea The indices $A$ and $B$ run over  the whole ten dimensional
spacetime.
\\ \par An interesting special case of the action (\ref{dbi1}) is
obtained when we restrict ourselves to purely radial fluctuations
of the D-brane in transverse space $\IR^4$ labelled by $\vec{x}$.
For such fluctuations the only excited field on brane is
$r(\xi_{\mu})$ and the angular variables remains fixed at their
initial values. This restriction on the  radial motion is
consistent since, for coincident NS5-branes, the background
(\ref{NS}) is $SO(4)$ invariant, and the D-brane experiences a
central force pulling it towards the origin.
\\ \par
Since the B-field (\ref{NS}) is in the angular directions, and the angular degrees of
freedom are not excited, the induced B-field in (\ref{dbi1}) vanishes. The DBI action
in this case is thus given by
 \be
 \label{dbi2} S=-T_p
\int d^{p+1}x {1\over \sqrt{H}}\sqrt{1+H\; \partial_{\mu}r \;
\partial^{\mu}r}\;. \ee
 As it was mentioned in \cite{Kutasov}, the form of the action (\ref{dbi2}) is very
 reminiscent of the DBI action of the tachyon in open string models \cite{tachyon}.
 After defining the variable $T$  the ``tachyon field" by
\be
\label{T}
T = \sqrt{k \alpha' +r^2} + \frac{1}{2} \sqrt{k\alpha'} \;{\mathrm ln}
\frac{\;\sqrt{k \alpha' +r^2}-\sqrt{k\alpha'\;}}{\;\sqrt{k \alpha' +r^2}+\sqrt{k\alpha'\;}}
\ee
the action (\ref{dbi1}) has the similar form of the tachyon action in open string models
\be
S_{tach} = - \int d^{p+1}x V(T) \sqrt{1+\partial_{\mu}T \; \partial^{\mu}T}, \;\;\;
\ee
where the asymptotic behavior of potential $V(T)$ is given by
\be
\label{pot}
{1\over T_p}\; V(T) \simeq \cases{{\mathrm exp}\;\frac{T}{\sqrt{k \alpha'}}  \;\;\;\ T
\rightarrow -\infty \cr\cr
1-\frac{k\alpha'}{2T^2}\;\;\;\;\;\;T \rightarrow \infty \cr}\;
\ee
Note that even when the D-brane is very close to the fivebranes, there is no perturbative
string tachyon here, because a fundamental string can not stretch between the D-brane and
the $NS5$-branes. \\
\par
In this case the ``tachyon field" $T$ acquires a geometrical
meaning and is related via (\ref{T}) to the distance between the
D-brane and the fivebranes. From the asymptotic behavior of the
potential (\ref{pot}) when $T\rightarrow -\infty$ we recognize  that
the potential goes exponentially to zero. This is precisely the
behavior exhibited at late times by the tachyon potential relevant
for rolling tachyon solution. This behavior of the potential leads
to the absence of plane wave solution around the minimum of the
potential at $T\rightarrow -\infty$, and to the exponential
decrease of the pressure at late times \cite{tachyonm}. For
unstable D-branes a similar equation of state is taken to signal
the decay of the D-brane into the closed string radiation, see
$e. g. $\cite{OkudaYD,llm, OkudaYD1}. In \cite{Kutasov} it was
proposed that in the case of D-brane in fivebranes background at
late times the D-brane decays into modes propagating in the fivebrane
throat. \\ \par In the next sections we restrict our attention to
the case of $p=3$, and we will try to investigate what types of
cosmologies arise from D3-branes rolling down the throat of NS5
brane.
\section{Coupling with gravity}
In \cite{Dccel}, motivated by the behavior of rolling scalar
fields in the strong coupling limit of theories, a new mechanism
for slow roll inflation was analyzed using the AdS/CFT
correspondence. The non-analytic behavior of the square-root in
the DBI action   gives rise to a speed limit restricting  how fast
the scalar field may roll. This is nothing but the Einstein's
causal speed limit in the holographic dimension, and is the
crucial bit of new physics in our inflationary mechanism. \\ \par
Motivated by this scenario, here we want to generalize the action
to incorporate the effects of coupling to four dimensional gravity
as well as to the other sectors suppressed by higher dimension
operators that may arise in the corresponding string
compactifications. We can also ask what is the low-energy
effective action of (\ref{dbi2}) when the gauge theory is coupled
to four dimensional gravity introducing by dynamical background
metric $g_{\mu\nu}$. A simple possible generalization can be given
by adding a potential term $V$ to the action that arise when the
system is coupled to four dimensional gravity and other sectors
involved in a full string compactification. Four dimensional
covariantization of (\ref{dbi2}) involving the potential $V$ is
\be \label{actiongravity} S = -{1\over \alpha' g_{YM}^2} \int d^4x
\sqrt{-g}\left({1\over \sqrt{f(\phi)}}\sqrt{1+f(\phi)\;
g^{\mu\nu}\partial_{\mu}\phi \partial_{\nu} \phi } + V(\phi)
\right) \ee In the expression above we have  use  field theory
variable which is defined by $\phi = r/\alpha'$, we have also
defined  $f(\phi) = \alpha'^2 + \lambda^2 /\phi^2$ and $\lambda^2
= \alpha' k$.
\\ \par
Now we can investigate some cosmological features of this system,
by studying FRW cosmologies solutions which follow from
(\ref{actiongravity}). We consider only flat cosmologies, \be
ds^2=-dt^2+a(t)^2dx^2. \nonumber \ee Since spatially inhomogeneous
terms are redshifted away during inflation, we consider the scalar
field ansatz $\phi = \phi(t)$. Here we also ignore the effects of
conformal coupling ${\mathcal R} \phi^2$, we will comment on the
reliability of this consideration further below and we will show
that its effects are self consistently negligible in our
solutions. With this ansatz the equation of motion can be
expressed by first defining $\gamma$ {\cite{Dccel}} analogous of
the Lorentz contraction factor in special relativity \be
\label{gamma} \gamma = \frac{1}{\sqrt{1-f(\phi)\dot{\phi}^2}} \ee
The energy density $\rho$ and pressure $p$ following from
(\ref{actiongravity}) are given by \bea \label{prho} \rho &=&
{\gamma \over \sqrt{f}} +V \cr\cr p&=& -\frac{1}{\gamma \sqrt{f}}
-V, \eea These definitions do not include overall coefficient of
$1/(\alpha' g_{YM}^2)$ which instead combines with $M_p$ in the
Einstein equations so that the scale $(M_p \sqrt{g_s})=M_p g_{YM}$
appears in all equations. The Friedmann equations read, \bea
\label{Friedmann} 3 H^2 &=& \frac{1}{g_s \alpha' M_p^2} \rho
\cr\cr 2 \frac{\ddot{a}}{a} + H^2 &=& - \frac{1}{g_s \alpha'
M_p^2} p \eea Here $H=\dot{a}/a$ is the usual Hubble parameter. We
will try to find solutions in which the brane is asymptotic to the
speed of light at late times in the gravity side background. This
means that the quantity $ \gamma$ grows when $t$ gets large, and
the behavior of $\rho$ will be substantially  different from the
usual case of $\rho=\frac{1}{2}\dot{\phi}^2 + V$, in which reduces
at small proper velocity. Firstly, we try to write the Friedmann
equations (\ref{Friedmann}) in a more suitable form. In fact they
can be integrated at once. It is referred to  the ``
Hamilton-Jacobi" formalism \cite{will}. Another perspective can be
obtained by viewing the resulting cosmology as a Wick rotation of
a BPS domain wall; the first order Friedmann equations are related
to the Bogolomoln'yi equations derived in \cite{dan}. \\ \par We
can derive the  Hamilton-Jacobi formalism equations by viewing the
scalar field $\phi$ as a time variable. In practice, this means
that we can consider $H = H(\phi)$ with $\phi= \phi(t)$. It is
important to notice that this assumption puts a limit on the
dynamics, since it considers $\phi$ as a  monotonic function, and
$H$ is a single valued function of $\phi$ and therefore it
restricts us to non-oscillatory behavior in $\phi$ \cite{Kutasov}.
\\ \par If we take time derivative of the first Friedmann equation
and then using the second equation, we find \be 6 H H' \dot{\phi}
= -\frac{1}{g_s\alpha' M_p^2} 3 H \gamma \sqrt{f} \dot{\phi}^2,
\ee which can be simply solved by using (\ref{gamma}) and it
results, \be \label{eq4} \dot{\phi} = \frac{-2H'}{\sqrt{f /
(g_s^2\alpha'^2 M_p^4) + 4 f H'^2}} . \ee Finally after
substituting into the first Friedmann equation and using
(\ref{prho}) we get an expression for the potential $V(\phi)$ in
terms of Hubble parameter $H(\phi)$, \be \label{eq3}
 V = 3 g_s\alpha'M_p^2 H^2 - \sqrt{(1+4 \; g_s^2 \alpha'^2 M_p^4 \; H'^2 )/f}
\ee Now if we are able to solve (\ref{eq3}) for a given potential
$V$, we can plug the solution into the expression (\ref{eq4}) to
find $\phi(t)$, and it can be used to find the scale factor as a
function of time. \\ \par Before trying to solve this equation,
let us discuss a little about the range of reliability of the
solutions. The DBI action is reliable to describe the dynamics of
D-brane, when the string coupling at the location of the brane is
small, $i. e.$ one must have exp$(\Phi) \ll 1$. Since from
equation (\ref{NS}) the dilaton grows without bound as one goes
down the throat of fivebranes, it gives us a lower bound on $\phi$
and also an upper bound on $g_s$
\be
g_s \ll 1
\;\;\;\;\;\;\;\;\;\; {\mathrm and} \;\;\;\;\;\;\;\;\;\;\;  \lambda
g_s/ \alpha' \ll \phi\;. \ee
In this limit, the DBI action is a good description for
D-brane dynamics, however there is also an upper limit for $\phi$.
In fact we are entitled to take $\phi$ as large as we wish
and for $\phi \gg M_p$, one finds standard slow roll expansion of
chaotic inflation. However, as a result of discussion in
\cite{Dccel}, models with $superPlanckian$ VEVs suffer
from destabilization from a slew of quantum corrections involving
for example gravitational conformal couplings $ {\mathcal R}
\phi^2$ \cite{dflation}. To avoid this, we shall instead restrict
ourselves to $subPlanckian$ VEVs, $\phi \ll M_p$. This allows us
to avoid destabilizing effects, circumventing some of the
difficulties involved in placing inflation within a string
comactification \cite{dflation, sw, tsey}. Combining these two condition we
find
\be
\label{limit}
g_s k / \lambda \ll  \phi \ll M_p
\ee
\par
In the next section we shall try to solve equation (\ref{eq3}) when $\phi$ lies in
(\ref{limit}) regime, for some relevant potentials.
\section{Cosmological Solution}
As mentioned before, we can consider the coupling of the system to
four dimensional gravity and also other sectors involved in full
string compactification. This whole could be described by a
potential $V(\phi)$ in action (\ref{actiongravity}). Consider
power series expansion of the potential $V(\phi)$ as \be V= V_0 +
V_2 \phi^2  + V_4 \phi^4 ... \ee The most important term in the
Lagrangian at low energy is the hard cosmological constant $V_0$.
As we will see below, giving a finite non-zero value to $V_0$,
dramatically affects    the solutions of (\ref{eq3}). This term
gets contribution in principle from all sectors of the system. In
effective field theory ( and approximately in string theory by
using the Bousso-Polchinski mechanism \cite{BP}) we may turn this
(close) to the value of interest for a given application. We will
make use of this freedom in our analysis.
\subsubsection*{case $V_0 \neq 0$}
Let us first consider the interesting case that will give rise to
inflation. We suppose that $V_0 \neq 0$ so that this term dominates
when $\phi$ is small. When $\phi$ lies in the regime
(\ref{limit}), the solution of equation (\ref{eq3}) can be
expanded as \be \label{H} H(T) = h_0 + h_1 \phi + h_2 \phi^2 + ...
\ee where coefficients $h_i$'s are \be h_0 = \sqrt{V_0 \over 3g_s
\alpha' M_p^2}, \;\;\;\;\;\;\; h_1= \frac{1}{\sqrt{12 \alpha' g_s
M_p^2 \lambda^2 V_0 - 4 g_s^2\alpha'^2 M_p^4}}\; ,\;\;\; ... \ee
In expression (\ref{H}) we can see that $V_{2i}$ dependence of
$h_j$ appears for ${j \geq 2i}$. Substituting this into the first
Friedmann equations, we find can $\phi(t)$ at late times and then from (\ref{H}) we
read scale factor as a function of time.
\\
\par
The condition for acceleration $(\ddot{a}>0)$ and D-cceleration
$(\gamma > 1)$ in terms of the potential $V$ can be written as
follows \footnote{I would like to thank E. Silverstein for her
useful comments and bringing my attention to this point.}
\cite{Dccel} \bea
{V^{3/2} \over M_p V' \phi} \;\sqrt {{3 k \over g_s}}\; &\gg& 1 \\
{g_s \over 3k V} \lambda^2 M_p^2 V'^2 &\gg& 1 \eea These are to be
contrasted with the usual slow roll conditions $(V'/V)^2M_p^2 \ll
1$ and $V''/V M_p^2 \ll 1$. \\ \par For example for a quadratic
potential of the form \be \label{p2} V=V_0 + V_2 \phi^2, \ee these
conditions can be satisfied for given $V_0 \sim {\mathcal
O}(g_sM_p^2)$ and $V_2 \sim {\mathcal O} (1/g_s)$. In this case it
is easy to see that the proper acceleration is much smaller than
string scale. This condition is necessary for self-consistency of
our solution.\\ \par The number of e-foldings is given by \be N_e
\sim \int_{\phi_f}^{\phi_i} \frac{d\phi}{M_p}\sqrt{\frac{k \; V}{3
g_s \phi^2}} \ee For the potential (\ref{p2}), it leads to $N_e
\sim \sqrt{k}/g_s$, which gives us sufficient large number in the
small $g_s$ limit.
\subsubsection*{case $V_0=0$ }
Let us now suppose that the constant term in the potential is
vanishing. As we will see in this case the solution of equation
(\ref{eq3}) has completely different behavior in comparison with
the the previous case with $V_0 \neq 0$. This comes from the fact
that when $\phi$ lies in the regime (\ref{limit}), the first term
in the Lagrangian (\ref{actiongravity}) will dominate and then we
expect a different behavior for the scale factor. Firstly, we want
to re-write the equation (\ref{eq3}) in a more tractable form. In
fact in this case, it is natural to write the equation in terms of
the ``tachyon field", which in the limit (\ref{limit}) is given by
 \be
 T \;\simeq\; \lambda \; {\mathrm ln}{\phi \over \lambda}
 \ee
Now in terms of ``tachyon filed", equation (\ref{eq3}) reads as \be
\label{eq7} V = 3 g_s M_p^2\;H^2 - \sqrt{e^{2 T/\lambda } \;+\; 4
g_s^2 \alpha'^2 M_p^4\;H'^2} \;, \ee while the condition
(\ref{limit}) restricts $T$ to the large negative values. Now the
equation (\ref{eq7}) can be solved approximately by \bea H(T) =
&-&\frac{2}{3 T}   \;-\; \frac{3 \lambda}{32g_s^2\alpha'^2M_p^4}\;
T^2 e^{2T/\lambda} \;+\; \frac{3 \lambda^2}{16g_s^2\alpha'^2M_p^4}
T e^{2T/\lambda} \cr\cr\cr
&+&\frac{\;(9+8V_2g_s\alpha'M_p^2)\lambda^3}{32g_s^2\alpha'^2M_p^4}
\; e^{2T/\lambda}
 \;\;+ ...
\eea substituting this into the expression (\ref{eq4}), we get the
equation of motion for $T$ at late times \be \dot{T} = -1 +
\frac{9T^4}{\;32 g_s^2M_p^4 \;}\;e^{2{ T /\lambda}} \; + ... \ee
Then the asymptotic behavior of T(t) for late times reads as \be
T(t)= - t + \frac{9}{64 \lambda g_s^2 \alpha'^2 M_p^4} \; t^4\;
e^{-2 t/\lambda} + ... \ee and we can determine the behavior of
the Hubble constant in this regime \be H(t)=\frac{2}{3t}+
\frac{3}{32 g_s^2M_p^4 \lambda}\;t^4e^{-2t/\lambda} + ... \ee
which looks like the matter dominant solution of the  Freidmann
equation, when the matter is a noninteractive dust. In fact for
any potential with $V_0=0$ and non-vanishing higher order terms,
the rolling scalar field looks to have equation of state $\omega
\approx 0$; which is a kinetic dust.
\section{Conclusion}
In this paper we have studied some cosmological solutions which
arise from D3-brane rolling down the throats of k coincident
NS5-branes. We started by coupling this system to four dimensional
gravity and studied possible FRW cosmological solutions. We have
shown that for various potentials $V$, which describe the coupling
of the system to four dimensional gravity (and also other sectors
involved in a full string compactification), different  solutions
appear. If the potential has a constant non-zero value in its
expansion we have inflationary solution for special range of
parameters. We also found that for $V_0=0$ the late times behavior
of the system is very similar to dynamics of tachyon in non-BPS
branes. In this case we found the late times behavior of scalar
field is similar to non-interactive dust. In the context of
tachyon condensation, the pressureless fluid behavior at late
times is usually a signal of the instability of the D-brane. This
has been argued to  provide a dual description of the closed
string radiation that D-brane decays into. As it was mentioned in
\cite{Kutasov} it is natural to assume that a similar
interpretation can be made here as well, with D-branes decaying into
modes living on the fivebrane.\\ \par For the inflationary
solution that we have found, it is interesting to analyze the
density perturbations resulting from inflationary solution found
in \cite{kflation}-\cite{juan}. The DBI Lagrangian is proportional
to $\sqrt{1-v_p^2}$, where $v_p= \lambda \dot{\phi}/\phi$ is the
gravity side proper velocity of the brane probe whose position
collective coordinate is the inflaton $\phi$. The inflationary
solution involves a proper velocity approach to speed of light as
$\phi$ approaches small values. Expanding the action in
fluctuations of $\phi$ involves expanding the square root in the
DBI Lagrangian, which produces powers of $\gamma=1/\sqrt{1-v_p^2}$
accompanying powers of the fluctuations of the inflation. Since
$\gamma$ is relatively large, this may produce a large
contribution to non-Gaussianities. This issue is under
investigation \cite{apear1}. \\ \par It is also interesting to
study D-brane dynamics in the deformed NS5-brane backgrounds and
cosmologies arising from that\cite{apear2}.
\section{acknowledgments}
I would like to thank S. Alexander, M.Alishahiha, A. Imaanpur, S. Parvizi, S. Sheikh-Jabbari
E. Silverstein for useful and interesting discussion.

\end{document}